\documentclass[%
 reprint,
 amsmath,amssymb,
 aps,
]{revtex4-2}

\usepackage{graphicx}
\usepackage{dcolumn}
\usepackage{bm}
\usepackage{bbold}
\usepackage{array}

\usepackage{physics}

\begin{document}

\preprint{APS/123-QED}

\title{Competition of vortex core structures in superfluid $^3$He-B}

\author{Riku Rantanen}
\author{Vladimir Eltsov}%
\affiliation{%
 Department of Applied Physics, Aalto University, PO Box 15100, FI-00076 AALTO, Finland
}%

\date{\today}

\begin{abstract}
Among vortex structures identified so far in superfluid $^3$He-B, the most common are the A-phase-core vortex and the double-core vortex. According to earlier numerical calculations, the double-core vortex is energetically favored nearly everywhere in the $p$-$T$ phase diagram. Nevertheless, in experiments the A-phase-core vortex has been observed down to temperatures of $0.6T_{\mathrm{c}}$ at high pressures. We use the Ginzburg-Landau formalism to calculate the energies of the two vortex structures in the experimentally relevant magnetic field as well as the energy barrier for the transition between the two structures. Assigning vanishing barrier as the boundary of the metastability region of the A-phase-core vortex, we reproduce the experimentally measured vortex phase diagram and provide an explanation for the reappearance of the double-core vortex near the critical temperature $T_{\mathrm{c}}$ at low pressures: The difference in Zeeman energy between the two vortex structures becomes relatively more important close to $T_{\mathrm{c}}$, and the A-phase-core vortex becomes unstable. In contrast to the equilibrium vortex structures, we suggest that the vortex nucleation process favors the A-phase-core vortex over the double-core vortex. Our approach can be used to analyze competition between different vortex structures in other unconventional superfluids and superconductors. 
\end{abstract}

\maketitle


\section{\label{sec:introduction}Introduction}
Due to macroscopic quantum nature, superfluids respond to rotation by forming topological defects, vortices with quantized circulation. In single-component superfluids such as $^4$He or in conventional superconductors, there exists only one type of quantized vortex, where the superfluidity is completely destroyed at the vortex axis and the phase of the order parameter winds by $2\pi$ around the vortex core. Such vortices are called singular. In contrast, the more complicated internal structure of the spin triplet p-wave superfluid $^3$He allows for a variety of different superfluid phases~\cite{VollhardtWolfle}, each supporting a number of vortex core structures~\cite{Salomaa1987}, which are non-singular, that is, superfluidity is preserved in the core. In such structures one can distinguish a hard core, within which the superfluid state differs from that in bulk and a soft core, where the state is the same as in bulk, but the orientation of the order parameter is different.

Existence of unconventional (non-singular or fractional) vortex structures is closely linked to a multicomponent order parameter~\cite{Babaev2007,Holmvall2023}. Besides superfluid $^3$He, such order parameters have been demonstrated or suggested in various systems including quantum gases and superconductors~\cite{Yao2016,Lagoudakis2009,Kang2019,Ishizu2022,Masuda2020,Xi2008,Yang2007,Zhao2020,Zhao2019,Cai2022,Seo2015,Seo2016,Chang2004,Ueda2022,Trautmann2018,Chomaz2023}. The search for multicomponent superfluidity and superconductivity is motivated in particular by the possibility of finding vortex structures with fractional vorticity, with applications in quantum information processing~\cite{Ivanov2001,Mizushima2024}. The recent discovery of potential spin-triplet superconductivity in UTe$_2$ with multiple superconducting phases~\cite{Aoki2020,Hayes2021,Aoki2022}, and the observation of anomalous vortex dynamics~\cite{Tokiwa2023} strongly suggest it has a multicomponent order parameter. To realize theoretically predicted unconventional vortices in experiments, it is important to understand competition between different structures which may form as local energy minima. Superfluid $^3$He provides a platform for studying competition between various non-singular and fractional~\cite{Autti2016} vortex structures in an environment where the order parameter form is explicitly known.

Already over 40 years ago, nuclear magnetic resonance (NMR) measurements in the rotating B phase of superfluid $^3$He revealed two distinct vortex types, one at low temperatures and pressures and another at high pressures closer to the A-B transition line~\cite{Ikkala1982,Hakonen1,Hakonen2,pekola1984,Krusius1984}. Later analysis~\cite{Salomaa1983,salomaa1985,Volovik1985,Thuneberg1986,Salomaa1986,Thuneberg1987} and measurements~\cite{Kondo1991} identified these two structures as the non-axisymmetric double-core vortex and the axisymmetric A-phase-core vortex, respectively, Figs.~\ref{fig:vortexstructures} and \ref{fig:softcore}. Pekola \textit{et al.}~\cite{pekola1984} measured the phase diagram of the transition between these two states and found a peculiar feature: at pressures around 17\,bar, double-core vortices were observed both at higher temperatures close to the superfluid transition temperature $T_{\rm c}$ and also at low temperatures, while at intermediate temperatures A-phase-core vortices were detected, Fig.~\ref{fig:phasediagram}. This re-entrant behavior has remained unexplained for decades. In this work, we numerically reproduce such shape of the experimentally measured transition line using the Ginzburg-Landau formalism, and give a qualitative explanation for the re-entrant behavior. In contrast to previous calculations, we are able to perform large scale simulations that include both the coherence length scale ($\xi \sim 10$\,nm) hard core and the dipole length scale ($\xi_d\sim 10\,$µm) soft core of the vortex in a single calculation. We find that the effect of the soft core on the vortex structure is crucial in understanding the observed phase diagram in a magnetic field.

An often overlooked feature of the original NMR measurements is the fact that they were done using a "start-stop" rotation scheme, where the sample was initially cooled to the lowest temperature and then rotation was successively started and stopped every 15 minutes during warming. This means that the sample was cleared of vortices and new ones nucleated at each measurement step. A simple comparison of vortex energies is then not enough to explain the observed phase diagram, as the lowest energy state is not necessarily the one that is nucleated. This can be seen for example in the A phase, where the hard-core single-quantum vortices have lower energy than soft-core double-quantum vortex skyrmions~\cite{DQVNature}, but are not nucleated when rotation is started below $T_{\rm c}$ as they have a higher critical velocity~\cite{Ruutu1997}. 

There are also measurements that were done with a full cooldown and warmup cycle in continuous rotation, with one example presented in Fig.~1 of Ref.~\cite{pekola1984}. This data shows that the vortex core transition has a strong hysteresis and indicates that the double-core vortex is actually the lowest-energy state in the majority of the $p$-$T$ plane of the B-phase region. Later numerical calculations have confirmed this conclusion~\cite{Thuneberg1987,saulsdiagram}. To reconcile this fact with the phase diagram observed in start-stop rotation, we suggest that the A-phase-core vortex has lower critical velocity and is nucleated when rotation is started.  Its metastability region, however, is limited in the $p$-$T$ plane. Beyond this region (that is, when the energy barrier separating two vortex structures vanishes), the A-phase-core vortex decays to the double-core vortex. We present here calculations of the energy barrier between the two structures as a function of pressure $p$ and temperature $T$ and demonstrate that the zero-barrier region agrees with the double-core-vortex region in the phase diagram of vortex states measured with the start-stop method, Fig.~\ref{fig:phasediagram}. 

Two vortex configurations similar to that of Fig.~\ref{fig:vortexstructures} have been recently proposed in the polar phase of spin-1 Bose-Einstein condensates \cite{Takeuchi2021b}. Our method of calculating the energy barrier could be used to predict transitions between different vortex structures in this and other systems where multiple distinct vortex types have been suggested~\cite{Takeuchi2021b,Liu2023,Zhang2017,Sauls2009,Tsutsumi2012,Borgh2017,Kobayashi2023}.

The paper is organized as follows. In Section~\ref{sec:GLtheory} we recap the Ginzburg-Landau formalism and in Section~\ref{sec:numerics} we present the numerical framework of our energy calculations. Section~\ref{sec:vortices} discusses the vortex structures of $^3$He-B and the effect of the spin-orbit interaction on their cores. The vortex core transition is discussed in Section~\ref{sec:transition}, along with the intrinsic magnetization of the vortices and the twisted double-core vortex state. Section~\ref{sec:summary} contains the summary and our conclusions. Appendix~\ref{app:coefficients} lists the coefficients used along with the strong-coupling corrections. Appendices~\ref{app:phases} and \ref{app:phasedetermination} list some of the relevant order parameter states seen in the vortex cores, and discuss a method of determining the nearest superfluid state for a given order parameter. Appendix~\ref{app:neb} gives a detailed introduction to the nudged elastic band method we use to calculate the energy barriers between the two vortex states.

\section{\label{sec:GLtheory}Ginzburg-Landau theory}
In the normal state above the critical temperature $T_\mathrm{c}$, helium-3 has the global phase symmetry $U(1)$, time-reversal symmetry $T$, and three-dimensional rotation symmetry in both spin and orbital spaces, $SO(3)_S$ and $SO(3)_L$. Below $T_\mathrm{c}$, some symmetries are broken in the superfluid transition. The degree of symmetry breaking is described in the Ginzburg-Landau theory by the order parameter, which for superfluid $^3$He with spin triplet p-wave pairing is a $3\times 3$ complex matrix $A_{\mu j}$. The first index $\mu$ corresponds to spin degrees of freedom and the second index $j$ to orbital ones.

The GL free energy functional $\mathcal{F}[A]$ consists of all independent terms that are invariant under the symmetry group of the normal fluid, up to fourth order in the order parameter. The bulk free energy density has one second order term and five fourth order terms \cite{Mermin1973}:
\begin{eqnarray}
    f_{\mathrm{bulk}}[A] = &&\alpha\Tr{AA^\dag} + \beta_1\vert\Tr{AA^T}\vert^2 \nonumber \\ &&+ \beta_2\left[\Tr{AA^\dag}\right]^2 + \beta_3\Tr{AA^T(AA^T)^*} \label{eq:fbulk} \\ &&+ \beta_4\Tr{(AA^\dag)^2} + \beta_5\Tr{AA^\dag(AA^\dag)^*} \nonumber
\end{eqnarray}
where $A^T$ is the transpose and $A^\dag$ the conjugate transpose of the order parameter. The coefficient $\alpha$ of the second-order term changes sign at the superfluid transition temperature $T_\mathrm{c}$, and controls the amplitude $\Delta\propto \sqrt{|\alpha|}$ of the order parameter below $T_\mathrm{c}$, while the $\beta_i$ parameters determine the lowest energy superfluid phase. The coefficients are discussed in more detail in Appendix~\ref{app:coefficients}.

In order to describe spatial variation of the order parameter, the free energy also includes terms quadratic in the gradients of the order parameter. There are a total of three gradient terms with coefficients $K_1$, $K_2$ and $K_3$:
\begin{eqnarray}
    f_{\mathrm{grad}}[A] = K_1(\nabla_k &&A^*_{\alpha j})(\nabla_k A_{\alpha j}) + K_2(\nabla_j A^*_{\alpha j})(\nabla_k A_{\alpha k}) \nonumber \\ &&+ K_3(\nabla_k A^*_{\alpha j})(\nabla_j A_{\alpha k}) 
    \label{eq:fgrad}
\end{eqnarray}
where summation over repeated indices is implied and $\nabla_i = \partial/\partial x_i$. The gradient energy favors a spatially uniform order parameter. Rotation of the whole system with an angular velocity $\Omega$ can be accounted for by replacing the gradient operators in Eq.~\eqref{eq:fgrad} with~\cite{Fujita1978}
\begin{equation}
    \nabla_k \rightarrow \mathcal{D}_k = \nabla_k - i\frac{2m_3}{\hbar}(\bm{\Omega}\times\bm{r})_k
    \label{eq:gradientD}
\end{equation}
where $\bm{r}$ is the relative position from the rotation axis $\bm{\hat{\Omega}}$, $2\pi\hbar/2m_3$ is the circulation quantum in superfluid $^3$He, and $m_3$ is the mass of the $^3$He atoms.

The bulk and gradient energies account for the majority of the total energy of the system. However, there are two additional terms that, while comparatively small in magnitude, turn out to be very important when considering the energy differences between competing minimum energy states. The first one is the Zeeman energy density in the presence of a magnetic field $\bm{H}$:
\begin{equation}
    f_{\mathrm{mag}}[A] = g_m H_i (AA^\dag)_{ij}H_j.
    \label{eq:fmag}
\end{equation}
An external magnetic field generally causes the order parameter components along the field to be suppressed. In bulk, this means that the isotropic B phase becomes less energetically favorable than the anisotropic A phase, which can reorient itself to avoid the suppression.

Finally, there is the spin-orbit interaction energy originating from the dipole interaction between the spins of the Cooper pair components depending on their relative orbital momentum. The dipole energy is roughly six orders of magnitude smaller than the bulk energy, but over long distances it is an important orienting force on the order parameter. The dipole energy density is
\begin{equation}
    f_{\mathrm{dip}}[A] = g_d\left[|\Tr{A}|^2 + \Tr{AA^*} - \frac{2}{3}\Tr{AA^\dag}\right].
    \label{eq:fdip}
\end{equation}
where $g_d$ is the coupling constant. While the dipole energy is too weak to cause big changes to the vortex core structure, it determines the size of the "soft core", a region where the order parameter recovers the minimum energy spin-orbit coupling state of the bulk. Through the soft core, the bulk dipole texture can influence the specific order parameter structure inside the hard core of the vortex. The soft cores are further discussed in Section~\ref{sec:softcores}.

The various coefficients entering the Ginzburg-Landau free energy functional define the lowest energy state of the system. Their values, listed in Appendix~\ref{app:coefficients}, are determined from the weak-coupling quasiclassical theory with strong-coupling corrections, and from experimental data. In the weak-coupling approximation, the lowest energy bulk phase in zero magnetic field is the B phase. Stabilization of the A phase requires strong-coupling corrections to the $\beta_i$ parameters. In zero magnetic field, the B phase has the order parameter structure
\begin{equation}
    A = e^{i\phi}\Delta_B R(\bm{\hat{n}},\theta)
    \label{eq:bphase}
\end{equation}
where $\Delta_B = \sqrt{|\alpha|/(6\beta_{12}+2\beta_{345})}$ with the summation convention $\beta_{ij..k} = \beta_i + \beta_j + ... + \beta_k$. The rotation matrix $R(\bm{\hat{n}},\theta)$ describes the relative orientation of the spin and orbital spaces as a rotation around the unit vector $\bm{\hat{n}}$ by the angle $\theta$.

With strong-coupling corrections, the A phase is stabilized at high pressures and temperatures. The order parameter in the A phase takes the form
\begin{equation}
    A_{\mu j} = \Delta_A\hat{d}_\mu (\hat{m}_j + i\hat{n}_j)
    \label{eq:aphase}
\end{equation}
where $\hat{\bm{d}}$ is the spin anisotropy vector, and the orthonormal orbital vectors $\hat{\bm{m}}$ and $\hat{\bm{n}}$ define the orbital angular momentum direction $\bm{\hat{l}} = \hat{\bm{m}}\cross\hat{\bm{n}}$. No explicit phase factor is written here, because a change of total phase is equivalent to the rotation of $\bm{\hat{m}}$ and $\bm{\hat{n}}$ around $\bm{\hat{l}}$. The order parameter amplitude is given by $\Delta_A = \sqrt{|\alpha|/4\beta_{245}}$. In contrast to the isotropic B phase, the A phase is anisotropic and has two gap nodes in the fermionic excitation spectrum along the axis defined by $\bm{\hat{l}}$. The anisotropy in spin space allows the bulk A phase to become stable in high magnetic fields, as the $\bm{\hat{d}}$ vector can reorient itself so as to avoid increase in the Zeeman energy. Inside a vortex core, however, the symmetries of the system can prevent this reorientation, as discussed further in Section~\ref{sec:aphasecore}.

The order parameter inside the cores of quantized vortices deviates from the bulk phase. We present the order parameter forms of other relevant states in Appendix~\ref{app:phases}. 

\section{\label{sec:numerics}Numerical methods}
For a given geometry and initial state, we search for the order parameter configuration that minimizes the Ginzburg-Landau free energy. In this work, we consider vortex structures in a cylindrical domain. For most of our calculations, we assume the system to be invariant in the $z$ direction and simulate only a two-dimensional disk.

In our simulation platform, the geometry is discretized into tetrahedral finite elements. This allows us to use geometries of any shape in three dimensions. The $3\times 3$ complex order parameter $A$ is defined at each vertex of the mesh, and the free energy functional is calculated in each tetrahedron and integrated over the whole system volume. To find the minimum energy state, we use the well established Limited-memory Broyden-Fletcher-Goldfarb-Shanno (L-BFGS) algorithm~\cite{Liu1989}. Our energy calculations take advantage of the parallel processing power of modern graphic processing units (GPUs), and we use the GPU compatible L-BFGS library CudaLBFGS~\cite{CudaLBFGS}.

For the two-dimensional disk simulations, our mesh consists of three circular layers stacked on top of each other. We set the boundary conditions in the $z$ direction to be periodic by copying the values of the bottom disk to the top disk. In order to stabilize the vortex inside our simulation box and to prevent it from escaping, we impose a discrete rotational symmetry condition on the outer boundary. We enforce a $\pi$ rotation symmetry condition by setting the order parameter at a boundary vertex $\bm{r}_1$ to be equal to the order parameter at the opposite boundary vertex $\bm{r}_2 = -\bm{r}_1$, rotated and phase shifted by $\pi$:
\begin{align}
    A_{\mu j}(\bm{r}_1) = e^{i\pi} R(\bm{\hat{z}},\pi)A_{\mu j}(\bm{r}_2)R^T(\bm{\hat{z}},\pi)
    \label{eq:C2}
\end{align}
where $R(\bm{\hat{z}},\pi)$ is a rotation around $\bm{\hat{z}}$ by $\pi$. This is done for all the vertices on the cylindrical boundary of the middle layer. Using this form of the boundary conditions enables us to include the soft core of the vortex in the calculations (see Section~\ref{sec:softcores}) without enforcing any definite form to the order parameter \textit{a priori}. Note that the boundary condition is only valid for a bulk B phase state with $\bm{\hat{n}} = \bm{\hat{z}}$, ie. far from the container walls and in an axial magnetic field.

Two distinct energy minima, for example the A-phase-core vortex and the double-core vortex states, can be separated from each other by an energy barrier. If the realized state is only metastable, the barrier can prevent it from transitioning to the true minimum. Calculating the barrier height requires finding a continuous minimum energy path (MEP) between the two states in the high dimensional state-space, which in our case for a mesh of $N$ vertices is an $18N$ dimensional real vector space. The MEP passes through a saddle point, which determines the height of the energy barrier.

We find the MEP by utilizing the nudged elastic band (NEB) method~\cite{Jonsson1998,Mills1995}, which has been widely used in chemical physics to find transition states between atom configurations~\cite{Jonsson2011,Asgeirsson2021} and in magnetic systems to calculate energy barriers between states~\cite{Dittrich2003,Bessarab2015}. We present a brief introduction to the method in Appendix~\ref{app:neb}.

\section{\label{sec:vortices}Vortex structures in the B phase}

\begin{figure}
    \centering
    \includegraphics[width=\linewidth,keepaspectratio]{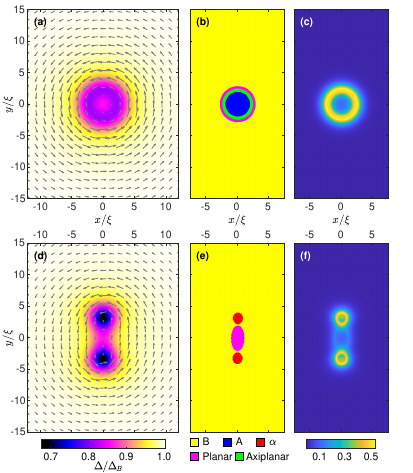}
    \caption{Stable vortex structures in superfluid $^3$He-B. \textbf{(a)-(c)} The axisymmetric A-phase-core vortex calculated at $p=25\text{ bar}$ and $T=0.87T_\mathrm{c}$. \textbf{(d)-(f)} The asymmetric double-core vortex calculated at $p=20\text{ bar}$ and $T=0.65T_\mathrm{c}$. Color in panels \textbf{(a)} and \textbf{(d)} shows the order parameter amplitude $\Delta = \Tr{AA^\dag}$, normalized to the bulk B phase value $\Delta_B$. The arrows indicate the superfluid velocity around the core, projected onto the xy-plane. Panels \textbf{(b)} and \textbf{(e)} show the closest superfluid phase at each point and colors in \textbf{(c)} and \textbf{(f)} show the distance from the nearest superfluid phase. The method to determine the phases is explained in Appendix~\ref{app:phasedetermination}.} 
    \label{fig:vortexstructures}
\end{figure}

When the system is rotated above some critical velocity, the irrotational superfluid B phase reduces the counterflow between the normal and superfluid components by forming vortices. Three possible vortex types in the B phase are suggested theoretically~\cite{Thuneberg1987,Salomaa1987}: the "o-vortex" with complete suppression of the superfluid state in the core, the symmetric A-phase-core vortex, and the asymmetric double-core vortex. Total suppression of the order parameter is energetically unfavorable, and the o-vortex state is never a global energy minimum. We focus our analysis on the two stable vortex states with finite order parameter in their cores, shown in Figure~\ref{fig:vortexstructures}. For clarity, the discussion first ignores spin-orbit coupling and its influence on the surrounding bulk and the details of the core structure. These effects will be included in Section~\ref{sec:softcores}. We also do not consider here spin-mass vortices \cite{Kondo19923331,Eltsov20004739}, since their formation requires strongly non-equilibrium dynamics.

\subsection{\label{sec:aphasecore}A-phase-core vortex}
The axially symmetric A-phase-core vortex, Fig.~\ref{fig:vortexstructures}(a)-(c), has been identified~\cite{Thuneberg1986,Salomaa1986} as the high temperature, high pressure structure observed in experiments~\cite{Ikkala1982,Hakonen1,Hakonen2,pekola1984}. The order parameter avoids the phase singularity by smoothly transitioning from the bulk B phase to the A phase. The symmetry of the superflow orients the orbital anisotropy vector $\bm{\hat{l}}$ of the A phase along the vortex axis $\bm{\hat{z}}$ at the center of the vortex. Orientation of the spin anisotropy vector $\bm{\hat{d}}$ is discussed in Section~\ref{sec:softcores}. For our case of bulk $\bm{\hat{n}} = \bm{\hat{z}}$, vector $\bm{\hat{d}}$ is also oriented along $\bm{\hat{z}}$. There is also a small component of the $\beta$ phase inside the core, which is responsible for the intrinsic magnetic moment of the vortex, also directed along $\bm{\hat{z}}$. 

At the center of the vortex, the order parameter has the form
\begin{equation}
    A(\bm{r}=0) = \begin{bmatrix}
        0 & 0 & b \\
        0 & 0 & ib \\
        a & ia & 0
    \end{bmatrix},
    \label{eq:Aphasecore}
\end{equation}
where $a$ and $b$ correspond to the A- and $\beta$-phase amplitudes, respectively, and $b$ is always smaller than $a$. We find the A-phase-core vortex state in our simulations by initializing the order parameter state with finite real parts for $A_{xz}$ and $A_{zx}$ and finite imaginary parts for $A_{yz}$ and $A_{zy}$ in the center, smoothly interpolating to the bulk B phase at the boundary with the appropriate $2\pi$ phase winding.

The transition from the bulk B phase to the A phase in the core proceeds through the planar and axiplanar states~\cite{Salomaa1987}, as shown in Fig.~\ref{fig:vortexstructures}(b). Approaching the core, the B phase becomes more and more planar-distorted all the way to the planar phase, where nodes in the energy gap appear along the azimuthal direction. Deeper in the core, the order parameter restructures itself from the planar phase to the A phase at the center. In terms of the axiplanar state described in Appendix~\ref{app:phases}, Eq.~\eqref{eq:axiplanarphase2}, this can be interpreted as $\varphi$ changing from $\pi/2$ to $0$, interpolating between the planar and A phases.

In zero magnetic field, the A-phase-core vortex is the lowest energy structure at higher pressures near the A-B phase transition temperature $T_{\mathrm{AB}}$~\cite{saulsdiagram}. This is natural as the energy difference between A and B phases decreases when the transition line is approached. The vortex remains metastable when cooled below the equilibrium temperature $T_{\mathrm{eq}}$, but decays into the asymmetric double-core vortex at the vortex transition temperature $T_{\mathrm{v}}$.

\subsection{\label{sec:doublecorevortex}Double-core vortex}

The non-axisymmetric double-core vortex appears in the vortex phase diagram at low temperatures and pressures~\cite{Kondo1991,Thuneberg1987,saulsdiagram}. The vortex carries a single quantum of circulation, but its core is split into two half cores, breaking the axial symmetry. The discrete symmetry of rotation by $\pi$ around the vortex axis still remains. Figure~\ref{fig:vortexstructures}(d) shows that the superfluid suppression is strongest inside the two half cores, and that the superfluid velocity around the vortex is non-axisymmetric close to the vortex. The core also generates superflow along the vortex axis~\cite{saulsdiagram}.

The broken axisymmetry allows to reduce the gradient energy (including the kinetic energy)~\cite{Thuneberg1986}. The bulk B phase smoothly transforms to the planar phase present between the two half cores, as shown in Figure~\ref{fig:vortexstructures}(e). The planar phase is oriented such that the gap suppression is in the y-direction (along the direction pointing from one half core to the other), and so is the spin-orbit axis $\bm{\hat{n}}$ (see Eq.~\eqref{eq:planarphase}). At the center of the vortex, the order parameter then has the form
\begin{equation}
    A(\bm{r}=0) = \begin{bmatrix}
        0 & 0 & -p \\
        0 & 0 & 0 \\
        p & 0 & 0
    \end{bmatrix},
    \label{eq:planarcore}
\end{equation}
where $p$ is the amplitude of the planar phase in the core.

Along the y-axis, the restructuring of the order parameter is more drastic on the path through the half cores. We find that the order parameter inside the two cores is closest to the $\alpha$ phase (see Eq.~\eqref{eq:alphaphase}), although in reality it is a more complicated non-unitary state as indicated by the intrinsic magnetization of the half cores, see Section~\ref{sec:magnetization}.

The double-core vortex is the lowest energy vortex structure in the majority of the phase diagram. We find the double-core vortex state in our simulations by taking the A-phase-core vortex as an initial condition and finding the minimum energy state in the region where it becomes unstable, for example at pressures below $15\text{ bar}$. The A-phase-core vortex decays to the double-core vortex, and we use the found state as an initial condition for further calculations in other regions of the phase diagram.

\subsection{\label{sec:softcores}Dipole energy and soft cores}

\begin{figure}
    \centering
    \includegraphics[width=\linewidth,keepaspectratio]{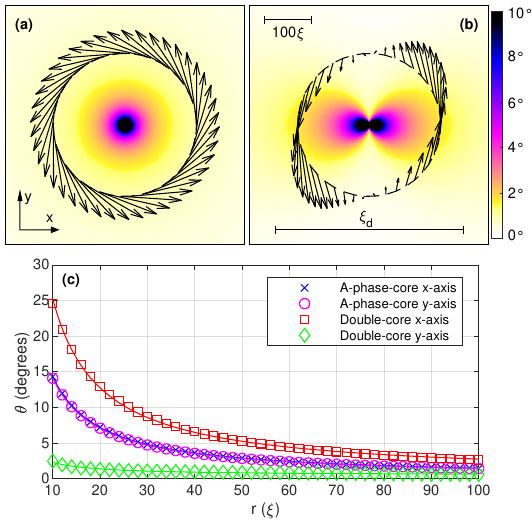}
    \caption{The soft cores of the symmetric A-phase-core vortex \textbf{(a)} and  the asymmetric double-core vortex \textbf{(b)} at $T = 1.8\text{ mK}$ and $p = 20$\,bar. The color indicates deviation $\theta$ from the bulk dipole angle $\theta_0$ and the arrows show $r\bm{\theta}$, which is independent of $r$ to first order. The half cores of the double-core vortex are positioned along the $y$ axis as in Fig.~\ref{fig:vortexstructures}. \textbf{(c)} The angle $\theta$ as a function of distance from the vortex core. The symbols show data from the simulations while the lines are fits to the model given in Ref.~\cite{softcore}, see Eq.~\eqref{eq:softcoremodel}.}
    \label{fig:softcore}
\end{figure}

The spin-orbit coupling in Eq.~\eqref{eq:fdip} affects the order parameter over distances of the dipole length $\xi_d\approx 10\text{ µm}$, which is much larger than the vortex core radius. In bulk, the rotation matrix $R(\bm{\hat{n}},\theta)$ obtains a uniform value on the scale of the dipole length. The value of $\theta = \theta_0$ in bulk is determined by minimizing the dipole energy \eqref{eq:fdip}, so that $\theta_0 = \cos^{-1}(-1/4) \approx 104.5^\circ$. The direction of the bulk $\bm{\hat{n}}_0$ vector is not defined by the dipole energy, but by the competition of the magnetic and surface energies. In a strong enough axial magnetic field the $\bm{\hat{n}}$ vectors form the flare-out texture, so that in the center of the cylinder $\bm{\hat{n}}$ is parallel to the field, and becomes tilted at the wall, so that the angle $\beta$ between the $\bm{\hat{n}}$ vector and the $z$-axis takes the value $\beta = \cos^{-1}(1/\sqrt{5}) \approx 63^\circ$.

The transition from the superfluid state in the core to the surrounding bulk B phase occurs over a length scale of the coherence length $\xi$, as determined by the competition of the gradient and condensation energies. Once the bulk phase is recovered, the orientation of the B-phase order parameter slowly recovers to the bulk values over a distance of $\xi_d$. The region $r < \xi_d$ is known as the soft core of the vortex. The rotation matrix of the B-phase order parameter in the soft core can be written as $R(\bm{\hat{n}},\theta) = R(\bm{\hat{n}}_0, \theta_0)R(\bm{\theta})$, where $\bm{\hat{n}}_0$ is the orientation in bulk and $\bm{\theta}$ defines the deviation from the bulk value by the angle $\theta = |\bm{\theta}|$ and the rotation axis $\bm{\hat{\theta}} = \bm{\theta}/\theta$. The profile of $R(\bm{\theta})$ in the soft core is independent of the bulk state $R(\bm{\hat{n}}_0,\theta_0)$, and is linked to the structure of the hard core, see Fig.~\ref{fig:softcore}. 

The structure of the hard core is mostly unaffected by the orientation of the order parameter in bulk, with the exception of the spin degrees of freedom. The slow spin rotation in the soft core links those in the hard core and in bulk. In particular, the $\bm{\hat{d}}$ vector in the A-phase-core vortex is fixed to $\bm{\hat{d}} = R(\bm{\hat{n}}_0,\theta_0)\bm{\hat{z}}$, so that the smooth soft core shown in Fig.~\ref{fig:softcore}a can connect the hard-core and bulk order parameters. Any other orientation will force the soft core to be asymmetric with $\bm{\theta}$ moving out of the transverse plane, which increases the gradient energy. We have confirmed this dependence of $\bm{\hat{d}}$ on $\bm{\hat{n}}_0$ by calculating the soft and hard cores of the A-phase-core vortex with different orientations of $\bm{\hat{n}}_0$. The orientation of $\bm{\hat{d}}$ is important to understand the energy of the vortex in an applied magnetic field~\cite{Kasamatsu2019}, as will be shown in Section~\ref{sec:transition}. The intrinsic magnetization, which is defined by the $\beta$ phase fraction of the core, is similarly rotated by the $R(\bm{\hat{n}}_0,\theta_0)$ matrix.

Due to the broken axisymmetry, the effect of the bulk spin-orbit rotation on the double-core vortex is more complicated. The two half cores of the vortex are aligned along the spin rotation axis $\bm{\hat{n}}$ of the planar phase between the cores (see Eq.~\eqref{eq:planarphase}), and this axis is oriented with respect to bulk $\bm{\hat{n}}_0$ in such a way that the deviation of $\theta$ in the hard core from the optimal angle $\theta_0$ is minimized~\cite{Thuneberg1987}. For $\bm{\hat{n}}_0=\bm{\hat{z}}$, the energy of the vortex does not depend on the orientation of the half cores.

In order to find the structure of the soft core, we simulate both the symmetric and asymmetric vortices in a cylindrical domain with the radius of $1000\xi$. We use a non-uniform grid resolution, with distance between grid vertices set to $0.1\xi$ within a distance $r < 10\xi$ from the vortex axis, and gradually increasing outside the core up to approximately $10\xi$ at the edge of the domain. This allows us to incorporate both the coherence-length-scale hard core and the dipole-length soft core in a single calculation. In these calculations, the size of the computational domain prevents the minimization process from removing the vortex from the system, so we have not applied the discrete rotational symmetry boundary condition described in Section~\ref{sec:numerics}. 

Due to the difference in energy scales between the bulk and dipole energies by about 6 orders of magnitude, the calculations must be done using double-precision arithmetics. Without boundaries, the orienting effect of the magnetic field on $\bm{\hat{n}}$ is too small to be observed even with double-precision calculations. To utilize the calculated soft core structures as initial conditions for further calculations, we initialize $\bm{\hat{n}}$ in the bulk to be along the vortex axis $\bm{\hat{z}}$, in order to preserve the discrete rotational symmetry. The dipole angle $\theta$ in the bulk is initially set to 90 degrees, and the minimization process eventually converges to the optimum $\theta = \theta_0$.

Laine and Thuneberg~\cite{softcore,spinwave} presented a model for $\bm{\theta}$ of an isolated vortex for distances $10\xi \lesssim r \ll \xi_d$:
\begin{align}
    \bm{\theta}(r,\phi) = &\frac{C_1\cos\phi}{r}\left(\frac{\sin\phi}{1+c}\bm{\hat{r}} + \cos\phi\bm{\hat{\phi}}\right) \nonumber \\
    &- \frac{C_2\sin\phi}{r}\left(\frac{\cos\phi}{1+c}\bm{\hat{r}} - \sin\phi\bm{\hat{\phi}}\right)
    \label{eq:softcoremodel}
\end{align}
where $r$ and $\phi$ are the radial and azimuthal coordinates in the two-dimensional plane perpendicular to the vortex axis. The coefficients $C_1$ and $C_2$ are determined by the vortex structure, so that for the axially symmetric A-phase-core vortex $C_1 = C_2$ and for the asymmetric double-core vortex $C_1/C_2 \gg 1$. 

The results of our soft core calculations are shown in Fig.~\ref{fig:softcore} for $T = 1.8\text{ mK} \approx 0.80T_\mathrm{c}$ at $p = 20$\,bar. The arrows in Fig.~\ref{fig:softcore}(a) and (b) visualize the orientation of $\bm{\theta}$ by plotting the radially independent value $r\bm{\theta}$. The orientation matches well the model in Eq.~\eqref{eq:softcoremodel}, for comparison see Fig.~1 in Ref.~\cite{spinwave}. Fig.~\ref{fig:softcore}(c) shows the radial dependence of the angle $\theta$ along the $x$ and $y$ directions for both vortices, in the range $10\xi \leq r \leq 100\xi$. The fits of the data to Eq.~\eqref{eq:softcoremodel} give the parameter values $C_1 = 44.7\xi$, $C_2 = 44.0\xi$ for the symmetric A-phase-core vortex and $C_1 = 78.1\xi$, $C_2 = 5.7\xi$ for the asymmetric double-core vortex. 

\begin{figure}
    \centering
    \includegraphics[width=\linewidth,keepaspectratio]{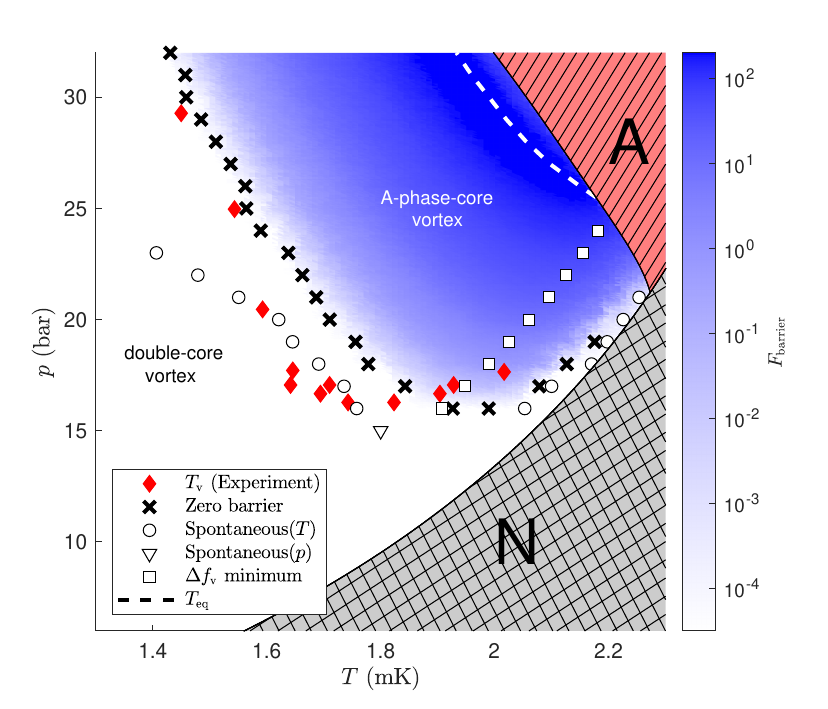}
    \caption{The vortex phase diagram in an axial magnetic field $H = 284$\,G. The color in the B phase region indicates the barrier height $F_{\text{barrier}}$ between the two states as defined in Eq.~\eqref{eq:barrierheight}. Experimental data from Ref.~\cite{Krusius1984} is marked with red diamonds. Crosses mark points from the simulations where the barrier disappears ($F_{\text{barrier}}=0$). The point of spontaneous transition in the simulation sweeps, ie.\ the point beyond which we could not stabilize the symmetric vortex, are marked with circles and triangles for temperature and pressure sweeps, respectively. Squares indicate the minimum of the normalized energy difference $\Delta f_{\mathrm{v}}$ taken from the curves in Fig.~\ref{fig:dfvsT}(a). The dashed line marks the equilibrium temperature $T_{\mathrm{v}}$ where the symmetric vortex becomes energetically favored, ie. the point where the lines in Fig.~\ref{fig:dfvsT}(a) cross zero.}
    \label{fig:phasediagram}
\end{figure}

\begin{figure*}
    \centering
    \includegraphics[keepaspectratio]{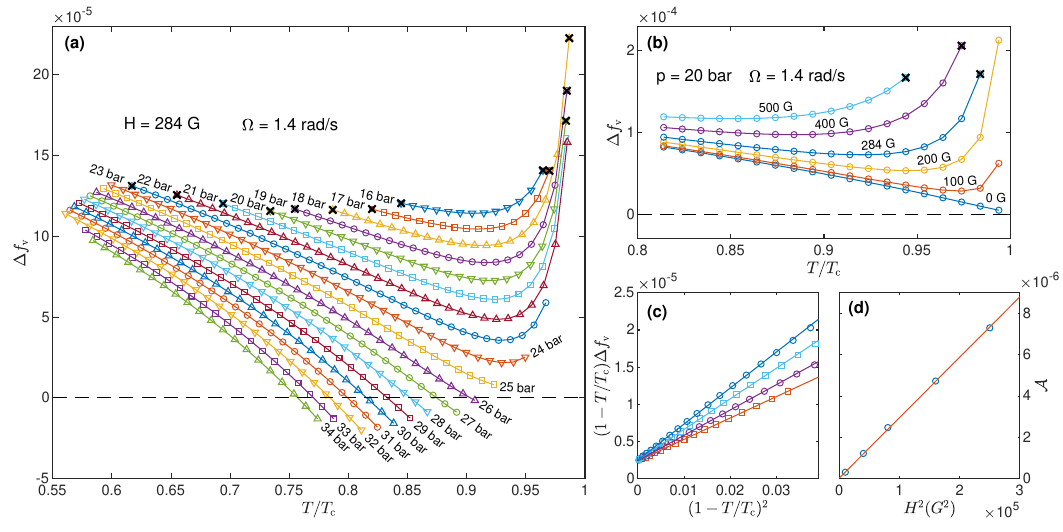}
    \caption{\textbf{(a)} The normalized energy density difference $\Delta f_{\mathrm{v}}$ between the two vortex states, Eq.~\eqref{eq:energydiff}, plotted as a function of temperature for different pressures. The magnetic field is $H = 284$\,G along the vortex axis. The crosses terminating some of the lines indicate a spontaneous vortex transition, beyond which we could not keep the A-phase-core vortex stable. The same transition is marked with circles in Fig.~\ref{fig:phasediagram}. The curves have a clear minimum as they approach the critical temperature, after which they diverge due to the difference in the Zeeman energy of the cores. \textbf{(b)} The normalized energy density difference as a function of temperature for different magnetic field strengths at $p = 20$\,bar. \textbf{(c)} Selection of data from (a), plotted in linearized coordinates (symbols matching panel (a)) and fit to Eq.~\eqref{eq:fitting} (lines). The constant term $\mathcal{C}$ has been substracted. From top down the lines correspond to $p = $23, 21, 19, and 17 bar. \textbf{(d)} The coefficient $\mathcal{A}$ for the fits in panel (c) as a function of $H^2$ (symbols) and a linear fit (line). The fit has a slope of $2.94\times 10^{-11}\text{ G}^{-2}$.}
    \label{fig:dfvsT}
\end{figure*}

\section{\label{sec:transition}Vortex core transition}
Experimentally, the vortex phase diagram of $^3$He-B has been measured using the "start-stop" rotation method during warmup~\cite{pekola1984,Krusius1984}. The system was cooled down to the lowest temperature and then allowed to warm up while rotating, stopping and reversing the rotation direction every 15 minutes. This means that for each data point, the previous vortices were completely removed from the system, and new ones nucleated. The vortex transition data from the experiment is marked in Fig.~\ref{fig:phasediagram} as red diamonds. The measurements were done in an axial magnetic field $H = 284$\,G and with angular velocity $\Omega = 1.4$\,rad/s.

The experimental data shows that above a temperature of approximately $0.6T_\mathrm{c}$ and above a pressure of 16\,bar, the symmetric A-phase-core vortex is nucleated when the system is rotated. Beyond the transition line, only the asymmetric double-core vortex is found. Near the critical temperature, the transition line curves upwards, showing the reappearance of the double-core vortex. 

Previous calculations~\cite{saulsdiagram} have shown that the double-core vortex is the lowest energy state in the majority of the phase diagram. The A-phase-core vortex only becomes favorable near the A-B transition line. From its appearance in the start-stop measurements~\cite{pekola1984} at low temperatures we conclude that the A-phase-core vortex is metastable, and easier to nucleate than the lower energy double-core vortex. Beyond the vortex transition line $T_\mathrm{v}$, we believe that the nucleated A-phase-core vortices become unstable and decay into double-core vortices.

To test our hypothesis, we find the minimum energy states of both vortex structures in the pressure and temperature range of the phase diagram in Fig.~\ref{fig:phasediagram}. The simulations are done using the same field and rotation values as in the experiment, $H = 284$\,G and $\Omega = 1.4$\,rad/s. As the initial states we use the results from the soft core calculations (Fig.~\ref{fig:softcore}), but we cut off the computational domain at $R = 30\xi$. We use the discrete rotational symmetry boundary condition described in Section~\ref{sec:numerics}. To obtain equilibrium states at each point of the phase diagram, we perform sweeps in both temperature and pressure, starting from the initial point $T = 1.8\text{ mK} \approx 0.80T_\mathrm{c}$ and $p = 20$\,bar. At each step of the sweep, the previous result is taken as the new initial condition. The temperature sweep step size is $0.01T_\mathrm{c}$ and the pressure sweep step size $1$\,bar.

At pressures below $24$\,bar, we were not able to stabilize the A-phase-core vortex in the full range of the temperature sweeps. At the points marked by circles in Fig.~\ref{fig:phasediagram}, the vortex spontaneously converts to the double-core vortex, indicating that the A-phase-core vortex is not a local minimum energy state at these temperatures. A similar instability occurs in the pressure sweep for $T = 1.8\text{ mK}$ at a pressure of $15$\,bar, marked by a triangle in Fig.~\ref{fig:phasediagram}.

The energy density difference between the two vortex states, shown in Fig.~\ref{fig:dfvsT}(a) and (b), is calculated as
\begin{equation}
    \Delta f_{\mathrm{v}} = \frac{F_{\text{A}}-F_{\text{D}}}{|f_B|V}
    \label{eq:energydiff}
\end{equation}
where $|f_B|$ is the bulk B phase energy density $f_B = \alpha\Delta_B^2/2$, $V$ is the container volume and $F_{\text{A}}$ and $F_{\text{D}}$ are the A-phase-core and double-core vortex energies, respectively. At low temperatures, the A-phase-core vortex is always higher energy, with the energy difference decreasing as either temperature or pressure are increased. However, around $T = 0.95T_\mathrm{c}$ the energy difference has a clear minimum, after which it starts increasing rapidly on approaching $T_\mathrm{c}$. At higher pressures, where the B phase doesn't extend up to the critical temperature, this effect is not visible, but instead the A-phase-core vortex becomes energetically favorable near $T_\mathrm{AB}$, although in a much smaller region than predicted at $H = 0$\,G~\cite{saulsdiagram}.

The divergence of the normalized energy difference near $T_\mathrm{c}$ only appears when a magnetic field is applied. Fig.~\ref{fig:dfvsT}(b) shows the energy difference $\Delta f_{\mathrm{v}}$ as a function of temperature for $p = 20$\,bar at different field strengths. The spontaneous transition of the A-phase-core vortex to the double-core vortex close to $T_\mathrm{c}$ is found for the field strengths $H = 284$\,G and above, and marked with black crosses in Fig.~\ref{fig:dfvsT} and circles in Fig.~\ref{fig:phasediagram}. 

The observed energy divergence can be understood by examining the temperature scaling of the difference in Zeeman energy (accounting for the normalization by $|f_B|$):
\begin{align}
    \Delta f_{\text{mag}} &= \frac{f_{\text{A}}^{\text{mag}}-f_{\text{D}}^{\text{mag}}}{|f_B|} \propto H^2\frac{g_m}{\alpha}\frac{\Delta^2}{\Delta_B^2} \nonumber \\ &\propto \frac{g_m}{N(0)}H^2 (1-T/T_\mathrm{c})^{-1}
\end{align}
which indicates that even a small difference in the Zeeman energy density in the cores of the two vortices can cause a divergence in the normalized energy difference close to $T = T_\mathrm{c}$, but only when $H\neq 0$. In the following, we present a qualitative reason for the difference in Zeeman energy densities between the two core structures, applicable at least in the axial magnetic field close to the center of the experiment sample, where $\bm{\hat{n}}_0 = \bm{\hat{z}}$. In this environment the order parameter in the center of the A-phase-core vortex has the form shown in Eq.~\eqref{eq:Aphasecore}, ie. with $\bm{\hat{d}}=\bm{\hat{z}}$. The corresponding magnetic energy density Eq.~\eqref{eq:fmag} is
\begin{equation}
    f_{\text{A}}^{\text{mag}} = g_m H^2 (AA^\dag)_{zz} = 2g_m a^2 H^2.
    \label{eq:fmagsym}
\end{equation}

In the double-core vortex, most of the magnetic energy density is concentrated in the planar phase region between the two half cores. The half cores themselves have lower Zeeman energy density than the surrounding bulk B phase. We can consider the planar core to have the order parameter form shown in Eq.~\eqref{eq:planarcore}, which gives the magnetic energy density
\begin{equation}
    f_{\text{D}}^{\text{mag}} = g_m H^2 (AA^\dag)_{zz} = g_m p^2 H^2
    \label{eq:fmagasym}
\end{equation}
where according to simulations $p\approx a$. The double-core vortex magnetic energy density is half of the A-phase-core vortex value Eq.~\eqref{eq:fmagsym}. The normalized magnetic energy difference between the two states is then finite, and diverges near $T_\mathrm{c}$ for a finite magnetic field.

Figures~\ref{fig:dfvsT}(c) and (d) show fits of the function
\begin{equation}
    f(1 - T/T_\mathrm{c}) = \mathcal{A}(1 - T/T_\mathrm{c})^{-1} + \mathcal{B}(1 - T/T_\mathrm{c}) + \mathcal{C}
    \label{eq:fitting}
\end{equation}
to the normalized energy difference curves. Our model predicts that $\mathcal{A}\propto H^2$. In the linearized coordinates of Fig.~\ref{fig:dfvsT}(c) it is clear that $\mathcal{A}$ is roughly pressure independent, and Fig.~\ref{fig:dfvsT}(d) shows the coefficient $\mathcal{A}$ as a function of $H^2$, taken from fits to the curves in Fig.~\ref{fig:dfvsT}(b). The fits support our hypothesis that the increase in $\Delta f_{\mathrm{v}}$ close to $T_{\mathrm{c}}$ is a direct consequence of the external magnetic field.

At first glance, the decrease of stability of the A-phase-core vortex in the magnetic field contradicts the increase of the stability of bulk A phase in the field. In bulk A phase, the $\bm{\hat{d}}$ vector is free to rotate and can be oriented such that $\bm{\hat{d}}\perp \bm{H}$, minimizing the Zeeman energy. In contrast, in the A-phase core the $\bm{\hat{d}}$ orientation is fixed by the soft core of the vortex and the surrounding B phase, as explained in Sec.~\ref{sec:softcores}.  If $\bm{H}$ points along this $\bm{\hat{d}}$ orientation, the Zeeman energy in the core is maximized. This applies to our calculations where $\bm{H}$, $\hat{\bm{n}}_0$ in bulk and $\hat{\bm{d}}$ in the core are all directed along $\hat{\bm{z}}$. The suppressing effect of the magnetic field on the stability of the A-phase-core vortex has been noted by Thuneberg~\cite{Thuneberg1987} and Kasamatsu et al.~\cite{Kasamatsu2019}, who studied the vortex structures at $T = T_c$.

\subsection{\label{sec:barrier}Energy barrier}

The relative increase in the energy of the A-phase-core vortex near $T_\mathrm{c}$ and spontaneous transition from the A-phase-core vortex to the double-core vortex on the upward temperature sweep in simulations supports the observation of the re-entrant behavior in the experimental phase diagram. To find more accurately the metastability region of the A-phase-core vortex, we use the nudged elastic band method described in Appendix~\ref{app:neb} to calculate the barrier height between two vortex states for each point in the phase diagram where we have calculated both structures.

The barrier height is indicated in Fig.~\ref{fig:phasediagram} by the background color in the B phase region. The barrier height is normalized to the difference in energies between the two states:
\begin{equation}
    F_{\mathrm{barrier}} = \frac{F_{\mathrm{max}} - F_{\mathrm{A}}}{|F_{\mathrm{A}} - F_{\mathrm{D}}|}
    \label{eq:barrierheight}
\end{equation}
where $F_{\mathrm{max}}$ is the energy of the highest energy state along the minimum energy path between the two vortex states.

The points where the barrier height becomes zero are indicated by black crosses in Fig.~\ref{fig:phasediagram}. They follow the behavior of the experimentally measured points, marked by red diamonds, with an excellent matching at high pressures and slightly shifted values at low pressures. Most importantly, we find that the energy barrier of the transition at low pressures becomes zero at both high and low temperatures and it is finite in between, replicating the re-entrant phase diagram observed in experiments. The re-entrant behavior is caused by the external magnetic field increasing the energy of the A-phase-core vortex, as is shown by the square markers in Fig.~\ref{fig:phasediagram}, corresponding to the minimums of the energy curves in Fig.~\ref{fig:dfvsT}(a). When the energy barrier disappears, the A-phase-core vortex becomes a saddle point in energy. In the calculated temperature sweeps, the conversion to the double-core vortex does not happen immediately, though, since the energy gradients around the saddle point are small.

Close to $T_{\mathrm{c}}$ the calculated metastability region of the A-phase-core vortex in Fig.~\ref{fig:phasediagram} is shifted towards higher temperatures when compared to the experimental data. The calculated values correspond to the completele disappearance of the energy barrier, while in the experiment the transition can be triggered by thermal fluctuations even when the barrier is finite. At low temperatures, the discrepancy between the experimental and calculated transition line may be caused by the inaccuracy of the strong-coupling corrections far from $T_{\mathrm{c}}$ and general limitations of the Ginzuburg-Landau model at low temperatures.

\subsection{\label{sec:magnetization}Vortex core magnetization}
\begin{figure}
    \centering
    \includegraphics[width=\linewidth,keepaspectratio]{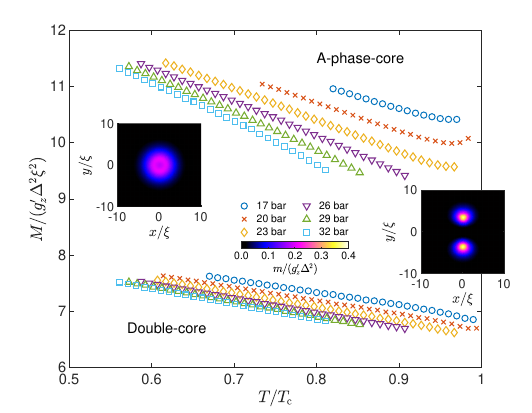}
    \caption{The intrinsic magnetization of the vortex cores as a function of temperature at different pressures. The markers on the top half of the figure correspond to the A-phase-core vortices and the bottom ones to the double-core vortex. The left and right insets show the magnetization density profiles for the A-phase-core and the double-core vortex, respectively.}
    \label{fig:magnetization}
\end{figure}

Both the A-phase-core and the double-core vortices have a finite intrinsic magnetization due to the non-unitary states present in their cores. The intrinsic magnetization density $\bm{m}$ can be expressed as
\begin{equation}
    m_{\kappa} = -ig'_{z}\epsilon_{\kappa\mu\nu}A_{\mu i}A_{\nu i}^*
    \label{eq:magnetization}
\end{equation}
which is non-zero only for non-unitary states, such as the $\beta$ phase found in the A-phase-core. The value of the coefficient $g'_z$ can be determined from the experimentally observed splitting of the A$_1$ phase transition temperature in high magnetic fields~\cite{Thuneberg1987}.

The total magnetic moment of the vortex is given by integrating Eq.~\eqref{eq:magnetization} over the vortex region. When the spin-orbit state of the bulk is taken into account (see Section~\ref{sec:softcores}), the magnetization is rotated by the bulk rotation matrix $R(\bm{\hat{n}}_0,\theta_0)$. Again for simplicity we only consider states with $\bm{\hat{n}}_0 = \bm{\hat{z}}$, which results in $\bm{M} = \int \bm{m}(\bm{r}) d\bm{r}$ pointing along the vortex axis.

We present the magnetic moments of both vortex states in Fig.~\ref{fig:magnetization}. In agreement with Thuneberg~\cite{Thuneberg1987}, we find that the A-phase-core vortex has approximately $50\%$ higher magnetization than the double-core vortex. The magnetization of the A-phase-core vortex shows slightly stronger pressure dependence than the double-core vortex.

The vortex magnetization can be linked to the experimentally measured gyromagnetic effect~\cite{Hakonen2}. The gyromagnetic effect is the difference in the NMR frequency shift for measurements done in magnetic fields oriented parallel to the rotation axis vs. antiparallel to the rotation axis. It is described by the textural energy term
\begin{equation}
    F_{\text{gm}} = \frac{4}{5}g_{\text{gm}}\kappa(\bm{H}\cdot(R\bm{\hat{z}}))
    \label{eq:Fgm}
\end{equation}
where the coefficient $\kappa$ can be directly determined from experimental measurements. The value of $\kappa$ is predicted to be directly proportional to the magnetization of the vortex cores~\cite{Thuneberg1987}:
\begin{equation}
    \kappa = \frac{5}{4}\frac{n_{\mathrm{v}}}{g_{\text{gm}}}M_z
\end{equation}
where $n_{\mathrm{v}}$ is the vortex density and $M_z$ the total magnetization of the vortex core, see Eq.~\eqref{eq:magnetization}.

The experimentally measured $\kappa$ has different behavior compared to the vortex magnetization calculated in numerical simulations. In experiments, the A-phase-core vortex has almost zero value for $\kappa$, while the double-core vortex has a much larger finite value. Our Ginzburg-Landau calculations indicate that the double-core vortex has a smaller magnetic moment than the A-phase-core vortex. The source of this discrepancy remains unclear, but explanations should probably be sought beyond the Ginzburg-Landau model, potentially in the effect of core-bound fermions. 

\subsection{\label{sec:twisted}Twisted double-core vortex}

\begin{figure}
    \centering
    \includegraphics{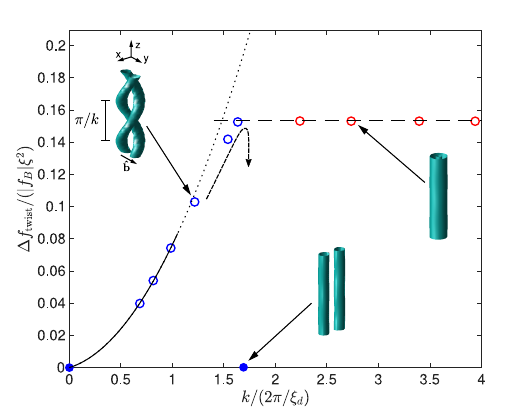}
    \caption{The energy difference per unit length along $z$ between twisted and untwisted double-core vortices $\Delta f_{\text{twist}}$, Eq.~\eqref{eq:ftwist}, as a function of the twist wave vector $k$ (symbols), with a quadratic fit (solid/dotted line) to the data points for $k < 1.2\times 2\pi/\xi_d$. The red and blue empty circles indicate A-phase-core vortices and twisted double-core vortices, respectively. Filled circles refer to the untwisted double-core vortex. The horizontal dashed line marks the energy difference between the A-phase-core vortex and the untwisted double-core vortex.
    The dashed arrow marks the transition from the twisted to untwisted double-core vortex on a gradual increase of $k$ in simulations. Inset illustrations show the isosurfaces for the order parameter amplitude $\Delta = \Tr{AA^\dag}$.}
    \label{fig:twisted}
\end{figure}

The double-core vortex can additionally break the translational symmetry along the vortex axis by twisting the two half cores around each other. This twisting has been created and observed in homogeneously precessing domain (HPD) NMR measurements~\cite{Kondo1991} and discussed theoretically~\cite{Krusius1993,softcore,spinwave}. The spin precession in the continuous-wave NMR creates a torque on the vortex core via the spin-orbit interaction and puts the core in the rotational motion. The vortex core orientation is assumed to be pinned by surface defects on the top and bottom of the container, which prevents uniform rotation of the whole vortex structure. Thus, two sub-cores of the double-core vortex twist around each other along the vortex axis. The twist of the hard core changes the structure of the soft core as well, which leads to observable NMR signatures \cite{softcore}. The broken translational symmetry also increases the total energy of the vortex due to additional gradients in the $z$ direction. The significant part of this energy is the kinetic energy associated with the axial supercurrents formed in the twisted state \cite{twisted}. We calculate the energy increase here.

The orientation of the double-core vortex in the plane perpendicular to the vortex axis can be defined by an anisotropy vector $\bm{\hat{b}}$, so that $\bm{\hat{b}}$ points from one half core to the other and 
\begin{equation}
    \bm{\hat{b}} = \cos\zeta \bm{\hat{x}} + \sin\zeta \bm{\hat{y}}
    \label{eq:bvector}
\end{equation}
where $\zeta(z) = kz$ is a function of $z$, and $k$ is the twist wave vector.

We calculate the energy of the twisted double-core vortex for different values of $k$ at a pressure of $20$\,bar and $T = 1.8\text{ mK}\approx 0.80T_c$. The computational domain is a three-dimensional cylinder created by stacking layers of circular disks with radius $R = 20\xi$ and distance between points in the disk approximately $0.2\xi$. For cylinders smaller than $256\xi$ in height, the number of layers is fixed at 400. For larger cylinders, the resolution of $256\xi/400 = 0.64\xi$ in the $z$ direction is kept fixed, and more layers are added to account for the increased size. The initial state for the minimization is constructed from the double-core vortex results in Section~\ref{sec:transition} by twisting the vortex along the cylinder axis, so that $\zeta$ goes from $0$ at the bottom layer to $2\pi$ at the top layer. The height of the simulation box is then $L = 2\pi/k$. The system has periodic boundary conditions on the top and bottom layers.

The difference in energy per unit length along $z$ between the twisted and untwisted double-core vortices is
\begin{equation}
    \Delta f_{\text{twist}} = \frac{F_{\text{twisted}} - F_{\text{untwisted}}}{L}
    \label{eq:ftwist}
\end{equation}
where $F_{\text{twisted}}$ and $F_{\text{untwisted}}$ are the total energies of the twisted and untwisted double-core vortices, respectively. Fig.~\ref{fig:twisted} shows $\Delta f_{\text{twist}}$ as a function of $k$. At smaller values of $k$, the energy difference is proportional to $k^2$, as is expected from the gradient energy, Eq.~\eqref{eq:fgrad}. However, at $k > 1.2\times2\pi/\xi_d$ the twisting starts to affect the core structure of the double-core vortex, squeezing the two half cores closer to each other. At this point the energy difference deviates from the quadratic behavior. The twisted double-core vortex state becomes energetically unfavored over the A-phase-core vortex around $k \approx 1.64\times2\pi/\xi_d$. More tightly wound states are found to be unstable, and convert to either the symmetric A-phase-core vortex or to the untwisted double-core vortex.

If the twisting is gradually increased starting from a state with $k < 1.64\times2\pi/\xi_d$ to $k$ values where double-core vortex loses stability, the minimization progresses through a nearly symmetric intermediate state and then to the untwisted double-core vortex. The A-phase-core vortex states shown in Fig.~\ref{fig:twisted} are found when the initial state is manually constructed at fixed $k$ by twisting the double-core vortex. In both cases after the twist is removed, also axial supercurrents related to the twist disappear in our simulations. 

The difference in NMR absorption between the twisted and untwisted states~\cite{Kondo1991} indicates that significantly lower values of $k \approx 0.21\times2\pi/\xi_d$~\cite{spinwave} are achieved in the HPD measurement. No conversion between vortex structures was observed in the experiment at those twist strengths, which is in agreement with our simulations. Simultaneously our findings show that the energetic preference of the double-core vortex is fragile to additional effects such as twisting. Especially in the highly nonequilibrium state after vortex nucleation it is conceivable that the A-phase-core vortex might exist in lower energy states than the double-core vortex, and we have shown that it is possible for the distorted double-core vortex to convert to the A-phase-core vortex. This may help to understand why the metastable A-phase-core vortex is the one that is nucleated in the experiments, while the double-core vortices appear when the A-phase-core state becomes unstable. However, one should be careful not to conflate vortex energetics with the nucleation process. Proper determination of which vortex state is nucleated requires analysis of their critical velocities, which can't be completely based only on the energy minimization.

\section{\label{sec:summary}Conclusion}
We have calculated the phase diagram of vortex structures in the B phase of superfluid $^3$He using the Ginzburg-Landau formalism. We reproduce the experimentally observed transition line between the axisymmetric A-phase-core vortex and the non-axisymmetric double-core vortex in a magnetic field oriented along the rotation axis. Good agreement is reached even though we considered only the case of the bulk $\bm{\hat{n}} $ vector oriented along the vortex axis. In experiments in a rotating cylinder this applies to vortices located near the cylinder axis. The re-entrant behavior of the transition at low pressures, where the double-core vortex re-appears close to the critical temperature $T_\mathrm{c}$, is found to originate from the increased energy of the A-phase-core vortex in the applied magnetic field, due to the unfavorable orientation of the $\bm{\hat{d}}$ vector along the magnetic field in the core. Overall agreement of the calculated phase diagram with experiments suggests that this phenomenon is more general than the particular case of $\bm{\hat{n}} $ orientation considered here. Based on this understanding, we expect that the phase diagram close to $T_{\rm c}$ depends on the direction of the magnetic field. In particular, if $\bm{\hat{d}}$ in the A-phase core is transverse to the applied field, we expect enhancement in the stability of the A-phase-core vortex, qualitatively similar to the phase diagram presented in Ref.~\cite{saulsdiagram}. The quantitative calculation of the effect of the bulk $\bm{\hat{n}} $ and of the magnetic field orientation remains a task for future research. 

Future simulations can also take into account the effect of walls, which is likely to shift the balance between the vortex structures as indicated by comparison of NMR and gyroscope measurements \cite{pekola1984}. It will be also interesting to obtain more experimental data close to $T_{\rm c}$ at low pressures, where our analysis predicts strong dependence of the transition line on pressure, and to measure the dependence of the transition line on the tilt angle of the magnetic field with the vortex axis and on the magnetic field strength.

The experimental data and our simulation results indicate that nucleation of A-phase-core vortices is preferred when the rotation is started in the superfluid state. This allows observation of A-phase-core vortices in the start-stop measurements in the region where they are only metastable. Double-core-vortices are found when the A-phase-core vortex becomes unstable and converts to the double-core vortex after nucleation. Further support of our hypothesis of the preferred nucleation of the A-phase-core vortices would require determining the critical velocities of the two vortex states. 


We use the nudged elastic band method to calculate the energy barrier between the two vortex states, and find that the barrier disappears in the region where the double-core vortex is seen in the experiments. This method of calculating the transition barrier can be used in the future to study other metastable states~\cite{Hindmarsh2024}, or to analyze the competition between different vortex structures in various unconventional superfluids and superconductors.

\begin{acknowledgments}
We thank Erkki Thuneberg for stimulating discussions. This work has been supported by the Academy of Finland (current name Research Council of Finland) with Grant No. 332964. We acknowledge the computational resources provided by the Aalto Science-IT project.

\end{acknowledgments}

\appendix

\section{\label{app:coefficients}Free-energy coefficients}
\begin{table*}
\caption{\label{tab:strongcoupling}%
Pressure dependence of the strong coupling corrections $\beta_i^{\text{sc}}$, the dipole energy coefficient $g_d$ and the spin-asymmetric Landau parameter $F_0^a$. The strong coupling corrections to the $\beta$ parameters are taken from Ref.~\cite{wimanthesis} in units of $|\beta_1^{\text{wc}}|$, the values of $g_d$ in units of $10^{31}\text{ erg}^{-1}\text{cm}^{-3}$ from Ref.~\cite{Thuneberg2001} and the values of $F_0^a$ from Ref.~\cite{F0a}.}
\begin{ruledtabular}
\begin{tabular}{cccccccc}
$p$(bar)&
$\beta_1^{\text{sc}}$&
$\beta_2^{\text{sc}}$&
$\beta_3^{\text{sc}}$&
$\beta_4^{\text{sc}}$&
$\beta_5^{\text{sc}}$& $g_d$&
$F_0^a$\\
\colrule
0.0 & -0.0020 & -0.0233 & -0.0133 & -0.0236 & -0.0759 & 2.7733 & -0.6986\\
2.0 & -0.0047 & -0.0397 & -0.0174 & -0.0323 & -0.1281 & 2.7950 & -0.7175\\
4.0 & -0.0070 & -0.0541 & -0.0215 & -0.0422 & -0.1736 & 2.8368 & -0.7282\\
6.0 & -0.0092 & -0.0671 & -0.0253 & -0.0528 & -0.2137 & 2.8987 & -0.7346\\
8.0 & -0.0112 & -0.0789 & -0.0288 & -0.0639 & -0.2494 & 2.9807 & -0.7389\\
10.0 & -0.0131 & -0.0899 & -0.0317 & -0.0753 & -0.2814 & 3.0829 & -0.7425\\
12.0 & -0.0150 & -0.1000 & -0.0341 & -0.0869 & -0.3103 & 3.2052 & -0.7459\\
14.0 & -0.0168 & -0.1096 & -0.0360 & -0.0989 & -0.3365 & 3.3476 & -0.7491\\
16.0 & -0.0186 & -0.1185 & -0.0373 & -0.1112 & -0.3601 & 3.5101 & -0.7520\\
18.0 & -0.0204 & -0.1270 & -0.0382 & -0.1242 & -0.3811 & 3.6927 & -0.7543\\
20.0 & -0.0222 & -0.1348 & -0.0387 & -0.1381 & -0.3995 & 3.8955 & -0.7559\\
22.0 & -0.0239 & -0.1421 & -0.0390 & -0.1530 & -0.4152 & 4.1184 & -0.7567\\
24.0 & -0.0255 & -0.1489 & -0.0390 & -0.1693 & -0.4279 & 4.3613 & -0.7569\\
26.0 & -0.0271 & -0.1550 & -0.0388 & -0.1872 & -0.4375 & 4.6245 & -0.7567\\
28.0 & -0.0285 & -0.1605 & -0.0386 & -0.2071 & -0.4438 & 4.9077 & -0.7563\\
30.0 & -0.0297 & -0.1655 & -0.0384 & -0.2291 & -0.4467 & 5.2111 & -0.7561\\
32.0 & -0.0308 & -0.1700 & -0.0381 & -0.2535 & -0.4462 & 5.5345 & -0.7560\\
34.0 & -0.0318 & -0.1742 & -0.0377 & -0.2804 & -0.4423 & 5.8781 & -0.7556\\
\end{tabular}
\end{ruledtabular}
\end{table*}

The $\alpha$ coefficient is a linear function of the temperature, and changes sign at $T_c$. It is given by
\begin{equation}
    \alpha(T) = \frac{1}{3}N(0)(1-T/T_c)
    \label{eq:alpha}
\end{equation}
with a pressure dependence given by the density of states for one spin direction $N(0) = m^* k_F/2\pi^2\hbar^2$, with $m^*(p)$ the effective mass and $k_F(p)$ the Fermi wavenumber. 

The $\beta_i$ parameters are derived in the weak-coupling theory as
\begin{align}
    -2\beta^{\mathrm{wc}}_1 = \beta^{\mathrm{wc}}_2 = \beta^{\mathrm{wc}}_3 = \beta^{\mathrm{wc}}_4 = -\beta^{\mathrm{wc}}_5 \nonumber \\ = \frac{7N(0)\zeta(3)}{240(\pi k_B T_c)^2}.
    \label{eq:betas}
\end{align}
We follow Wiman and Sauls \cite{saulschannels} and use strong-coupling corrections to the $\beta_i$ parameters in the form of
\begin{equation}
    \beta_i(p,T) = \beta^{\mathrm{wc}}_i(p) + \frac{T}{T_c}\beta^{\mathrm{sc}}_i(p)
    \label{eq:strongcoupling}
\end{equation}
where $\beta^{\mathrm{sc}}_i$ are the strong-coupling corrections. We use the $\beta_i^{\mathrm{sc}}$ values from Ref.~\cite{wimanthesis}. These values along with the given temperature dependence account reasonably well for the bulk A-B transition line within the Ginzburg-Landau theory. As Ref.~\cite{wimanthesis} includes only a plot, we have tabulated the used values in Table~\ref{tab:strongcoupling}.

The gradient energy coefficients $K_1$, $K_2$ and $K_3$ have the same pressure-dependent value in the weak-coupling theory, given by
\begin{equation}
    K \equiv K_1 = K_2 = K_3 = \frac{7\zeta(3)}{60}N(0)\xi_0^2
    \label{eq:K}
\end{equation}
where $\xi_0 = \hbar v_F/2\pi k_B T_c$ is the zero-temperature coherence length. The value of $K$ determines the relevant length scales of the problem. For instance, any localized suppression of the superfluid density (such as in the vortex cores) recovers smoothly to the bulk value within a distance of the Ginzburg-Landau coherence length $\xi = \sqrt{K/|\alpha|} \approx 10\text{-}100\text{ nm}$ depending on pressure and temperature. Similarly deviations from the minimum energy spin-orbit coupling state of the bulk liquid, determined by the dipole energy \eqref{eq:fdip}, are smoothed out over distances of the dipole length $\xi_d = \sqrt{K/g_d} \approx 10\text{ µm}$.

The Zeeman energy coefficient $g_m$ defines the strength of the magnetic field interaction:
\begin{equation}
    g_m = \frac{7\zeta(3)}{48\pi^2}\frac{N(0)(\gamma\hbar)^2}{[(1+F_0^a)k_BT_c]^2}
    \label{eq:gm}
\end{equation}
where $\gamma = -20 378\text{ G}^{-1}\text{s}^{-1}$ is the gyromagnetic ratio of helium-3 and $F_0^a$ is the pressure dependent spin-asymmetric Landau parameter~\cite{F0a} with values tabulated in Table~\ref{tab:strongcoupling}.

The spin-orbit coupling coefficient $g_d$ has not been accurately determined from theoretical considerations, so we use values extracted from experiments~\cite{Thuneberg2001}. The dipole coefficient is approximately temperature-independent, and its value (tabulated in Table~\ref{tab:strongcoupling}) ranges from $2.8\times 10^{31} \text{ erg}^{-1}\text{cm}^{-3}$ at zero pressure to $5.9\times 10^{31} \text{ erg}^{-1}\text{cm}^{-3}$ at $p = 34\text{ bar}$.

\section{\label{app:phases}Superfluid phases}
Superfluid helium-3 has a wide variety of possible phases and states. The vortex cores in $^3$He-B contain a few other states in addition to the A (Eq.~\eqref{eq:aphase}) and B (Eq.~\eqref{eq:bphase}) phases described in the main text. We give the order parameter forms of the relevant states in this Appendix.

The planar phase is never stable in the bulk. It can, however, be found between the two half cores of the asymmetric double-core vortex. It is also an important transitional state between the A and B phases, for example in the A-phase-core vortex. Its order parameter structure is
\begin{equation}
    A = e^{i\phi}\Delta_{\mathrm{pl}} R(\bm{\hat{n}},\theta)(\mathbb{1} - \bm{\hat{w}}\bm{\hat{w}}^T)
    \label{eq:planarphase}
\end{equation}
where $R$ is a rotation matrix like in the B phase, $\mathbb{1}$ is the identity matrix and $\bm{\hat{w}}$ is a unit vector that defines the direction of the gap suppression. The amplitude is $\Delta_{\mathrm{pl}} = \sqrt{|\alpha|/(4\beta_{12} + 2\beta_{345})}$. The planar phase in magnetic field can also avoid increases in the Zeeman energy by reorienting $\bm{\hat{w}}$, but like in the A phase, the vortex configuration can prevent this effect.

Recently, the $\beta$ phase has been observed in confined helium \cite{Dmitriev2021}. It is also known to be found in small proportions in the symmetric A-phase-core vortex in the B phase. Its order parameter is similar to the A phase, with the orbital and spin degrees of freedom swapped:
\begin{equation}
    A_{\mu j} = \Delta_\beta (\bm{\hat{d}}_\mu + i\bm{\hat{e}}_\mu)\bm{\hat{l}}_j
    \label{eq:betaphase}
\end{equation}
where $\bm{\hat{d}}$ and $\bm{\hat{e}}$ are orthonormal vectors in spin space, and $\bm{\hat{l}}$ is the orbital anisotropy direction. The amplitude is $\Delta_\beta = \sqrt{|\alpha|/4\beta_{234}}$.

In our simulations of the vortex core structures, we find two more states, the $\alpha$ state in the half-cores of the double-core vortex and the axiplanar state in the transitional region between the bulk B phase and the A phase core of the symmetric vortex. An example of the $\alpha$ phase order parameter is
\begin{equation}
    A = e^{i\phi}\Delta_\alpha\begin{bmatrix}
        1 & 0 & 0 \\
        0 & e^{i\pi/3} & 0 \\
        0 & 0 & e^{-i\pi/3}
    \end{bmatrix}
    \label{eq:alphaphase}
\end{equation}
and the amplitude $\Delta_\alpha = \sqrt{|\alpha|/(6\beta_2 + 2\beta_{345})}$.

The axiplanar state is an equal spin pairing state with each spin projection having different orbital momenta. Generally it can be written as
\begin{align}
    A_{\mu j} = &\frac{1}{2}\Delta_\uparrow(\bm{\hat{d}} + i\bm{\hat{e}})_\mu(\bm{\hat{m}}_\uparrow + i\bm{\hat{n}}_\uparrow)_j \nonumber \\ + &\frac{1}{2}\Delta_\downarrow(\bm{\hat{d}} - i\bm{\hat{e}})_\mu(\bm{\hat{m}}_\downarrow + i\bm{\hat{n}}_\downarrow)_j
    \label{eq:axiplanarphase1}
\end{align}
where the $\uparrow$ and $\downarrow$ indices indicate the variables corresponding to spin-up and spin-down populations with $\bm{\hat{l}}_\uparrow = \bm{\hat{m}}_\uparrow \times \bm{\hat{n}}_\uparrow$ and $\bm{\hat{l}}_\downarrow = \bm{\hat{m}}_\downarrow \times \bm{\hat{n}}_\downarrow$. In zero magnetic field the spin populations are equal and $\Delta_\uparrow = \Delta_\downarrow$ The axiplanar state contains both the planar and A phase states as special cases, when $\bm{\hat{l}}_\uparrow = -\bm{\hat{l}}_\downarrow$ and $\bm{\hat{l}}_\uparrow = \bm{\hat{l}}_\downarrow$, respectively. When interpolating between these two states (as in the A-phase-core vortex), we can restrict the order parameter to the form
\begin{align}
    A_{\mu j} = &\Delta_{ap}(\bm{\hat{d}}+i\bm{\hat{e}})_\mu\left[\bm{\hat{m}} + i(\bm{\hat{n}}\cos\varphi + \bm{\hat{l}}\sin\varphi)\right]_j \nonumber \\
    + &\Delta_{ap}(\bm{\hat{d}} - i\bm{\hat{e}})_\mu\left[\bm{\hat{m}} + i(\bm{\hat{n}}\cos\varphi - \bm{\hat{l}}\sin\varphi) \right]_j
    \label{eq:axiplanarphase2}
\end{align}
where $2\varphi$ is the angle between the $\bm{\hat{l}}_\uparrow$ and $\bm{\hat{l}}_\downarrow$ vectors. Then  $\varphi = 0$ gives the A phase Eq.~\eqref{eq:aphase} and $\varphi = \pi/2$ gives the planar phase Eq.~\eqref{eq:planarphase}.

\section{\label{app:phasedetermination}Determination of the nearest phase}

\begin{table}[h]
\caption{\label{tab:inertphases}%
The signature values of the fourth order bulk energy terms for each phase.}
\begin{ruledtabular}
\begin{tabular}{cccccc}
\textrm{Phase}&
$I_1$&
$I_2$&
$I_3$&
$I_4$&
$I_5$\\
\colrule
\textrm{B} & 1 & 1 & 1/3 & 1/3 & 1/3 \\
\textrm{Planar} & 1 & 1 & 1/2 & 1/2 & 1/2 \\
\textrm{Polar} & 1 & 1 & 1 & 1 & 1 \\
$\alpha$ & 0 & 1 & 1/3 & 1/3 & 1/3 \\
\textrm{Bipolar} & 0 & 1 & 1/2 & 1/2 & 1/2 \\
\textrm{A} & 0 & 1 & 0 & 1 & 1 \\
$\beta$ & 0 & 1 & 1 & 1 & 0 \\
$\gamma$ & 0 & 1 & 0 & 1 & 0 \\
\end{tabular}
\end{ruledtabular}
\end{table}

We follow the classification scheme for inert phases by Barton and Moore~\cite{Barton1974}. The fourth order terms in Eq.~\eqref{eq:fbulk} for a normalized order parameter $\Tilde{A} = A/\sqrt{\Tr{AA^\dag}}$ can be written as $\sum_{i=1}^5 \beta_i I_i$, where the values of $I_i$ are invariant for each inert phase and act as their signatures. The signatures of each phase are listed in Table~\ref{tab:inertphases}. Note that the axiplanar state is not included, since it is non-inert, ie. the values $I_i$ are not invariant.

Using these signature values, we determine the closest inert phase for an arbitrary order parameter $\Tilde{A}$ by calculating the values of the fourth order terms and comparing them to those of the inert phases listed in Table~\ref{tab:inertphases}. For example, the distance from the B phase is given by
\begin{equation}
    D_B = \sqrt{\sum_{i=1}^5 (I_i[\Tilde{A}] - I_i^B)^2}
    \label{eq:inertdistance}
\end{equation}
where $I_i^B$ are the values of $I_i$ for the B phase. The closest phase is one that minimizes the distance $D$.

This only gives a rough determination of the closest inert phase. Actually in the cores of vortices in the B phase, the order parameter can be in non-inert states, such as the axiplanar state in the A-phase-core vortex, see Fig.~\ref{fig:vortexstructures}(b). The closest inert phase is the bipolar phase which has no physical relevance here. Similarly in the half cores of the double-core vortex (see Fig.~\ref{fig:vortexstructures}(e), the order parameter is closest to the $\alpha$ phase, but in reality is in a more complicated non-unitary state. This can be seen for example from the intrinsic magnetization of the half cores shown in Fig.~\ref{fig:magnetization}, which should be zero for a pure $\alpha$ phase order parameter.

\section{\label{app:neb}Nudged elastic band method}

We represent the system configuration as a real vector $\bm{X}^i$ in an $18N$ dimensional space defined as
\begin{equation}
    \bm{X}^i = \left[\bm{A}^1, \bm{A}^2, ..., \bm{A}^N\right]
    \label{eq:Rspace}
\end{equation}
where the order parameter vector $\bm{A}^j$ at node $j$ is
\begin{equation}
    \bm{A}^j = \left[\Re{A_{xx}^j},\Im{A_{xx}^j},\Re{A_{xy}^j}, ..., \Im{A_{zz}^j}\right]
\end{equation}
The NEB method starts with an initial discretized path consisting of $Q-2$ intermediate states $\bm{X}^i$ between the two end points $\bm{X}^1$ and $\bm{X}^Q$. The end points are kept fixed while the intermediate states are iteratively adjusted ("nudged") so that they converge towards the minimum energy path. The direction of the nudges is determined by the functional gradient $-\nabla \mathcal{F}$, with the observation that only adjustments transverse to the path should be included. Movement along the path only changes the local resolution of our discretization. The transverse component is given by
\begin{equation}
    \nabla \mathcal{F}^i|_\perp = \nabla \mathcal{F}^i - (\nabla \mathcal{F}^i\cdot\bm{\hat{\tau}}^i)\bm{\hat{\tau}}^i
    \label{eq:transverseforce}
\end{equation}
where $\mathcal{F}^i = \mathcal{F}(\bm{X}^i)$ and $\bm{\hat{\tau}}^i$ is the unit tangent direction along the path at state $i$. Determining the tangent vector accurately is crucial for the performance of the method.

Due to the finite number of intermediate states on the path, this iteration alone can cause them to accumulate at valleys of the energy landscape. We want to avoid this, as we are specifically interested in the maximum along the path. To keep a good enough resolution throughout, we introduce a spring force on the states that acts to keep them equally separated. The spring force should only act in the direction parallel to the path, so as not to interfere with the minimization effort. The force is given by
\begin{equation}
    \bm{F}_s^i|_{||} = \kappa\left[|\bm{X}^{i+1} - \bm{X}^i| - |\bm{X}^i - \bm{X}^{i-1}|\right]\bm{\hat{\tau}}^i
    \label{eq:springforce}
\end{equation}
where $|\bm{X}^{i+1}-\bm{X}^i|$ is the Euclidean distance between states $i$ and $i+1$ in the $18N$ dimensional space. The total NEB force is then
\begin{equation}
    \bm{F}_{\text{NEB}}^i = \nabla \mathcal{F}^i|_\perp + \bm{F}_s^i|_{||}
    \label{eq:fneb}
\end{equation}

A poor estimate of the path tangent $\bm{\hat{\tau}}^i$ can lead to instabilities or slow convergence. We use a definition that uses either forward or backward differences depending on the energy of the state and its neighbours~\cite{Henkelman2000}, so that
\begin{equation}
    \bm{\tau}^i = 
    \begin{cases}
        \bm{\tau}_+^i = \bm{R}^{i+1}-\bm{R}^i, & \text{if } \mathcal{F}^{i+1} > \mathcal{F}^i > \mathcal{F}^{i-1} \\
        \bm{\tau}_-^i = \bm{R}^{i}-\bm{R}^{i-1}, & \text{if } \mathcal{F}^{i+1} < \mathcal{F}^i < \mathcal{F}^{i-1}
    \end{cases}.
\end{equation}
In the case where the intermediate state $i$ is higher or lower in energy than both neighboring states, we take the weighted average of $\bm{\tau}_+^i$ and $\bm{\tau}_-^i$, with the weights determined by the energy differences:
\begin{equation}
    \bm{\tau}^i =
    \begin{cases}
        \bm{\tau}_+^i\Delta \mathcal{F}_{\max}^i + \bm{\tau}_-^i\Delta \mathcal{F}_{\min}^i, & \text{if } \mathcal{F}^{i+1} > \mathcal{F}^{i-1} \\
        \bm{\tau}_+^i\Delta \mathcal{F}_{\min}^i + \bm{\tau}_-^i\Delta \mathcal{F}_{\max}^i, & \text{if } \mathcal{F}^{i+1} < \mathcal{F}^{i-1}
    \end{cases}
\end{equation}
where $\Delta \mathcal{F}_{\max}^i = \max(|\mathcal{F}^{i+1} - \mathcal{F}^i|, |\mathcal{F}^{i-1} - \mathcal{F}^i|)$ and $\Delta \mathcal{F}_{\min}^i = \min(|\mathcal{F}^{i+1}-\mathcal{F}^i|, |\mathcal{F}^{i-1}-\mathcal{F}^i|)$. Note that $\bm{\tau}^i$ is normalized to a unit vector in Eqs.~\eqref{eq:transverseforce} and \eqref{eq:springforce}.

To find the minimum energy path, we use the Global L-BFGS method as described in Ref.~\cite{globalLBFGS}. In this method, for each step of the iteration, the step is calculated for the whole path $\bm{x} = [\bm{X}^1, ..., \bm{X}^Q]$ using the combined NEB force $\bm{F} = [\bm{F}_{\text{NEB}}^1, ..., \bm{F}_{\text{NEB}}^Q]$ as the search direction. This is in contrast to calculating a step for each intermediate state separately, in which case the approximated Hessian would not contain any information on interactions between the states on the path. We use the version of the method without a line search, as it is computationally much less expensive and convergences rapidly enough for good estimations of the initial Hessian.


\bibliography{phasediagram}

\end{document}